\documentclass[twocolumn, pra, aps,showkeys]{revtex4-1}

 \pdfoutput=1

\usepackage{amsmath, amssymb, amsthm}
\usepackage{graphicx}
\usepackage{graphics}
\usepackage{color}
\usepackage{subfigure}
\usepackage{epstopdf}
\usepackage{ifthen}

\newcommand{\REM}[1]{\ifthenelse{0=1}{{#1}}{}}
\newcommand{\mc}[1]{\mathcal{{#1}}}
\newcommand{\wb}[1]{\overline{{#1}}}
\newcommand{\etal}{\textit{ et al. }}
\newcommand{\be}{\begin{equation}}
\newcommand{\ee}{\end{equation}}
\newcommand{\ba}{\begin{array}}
\newcommand{\ea}{\end{array}}
\newcommand{\bi}{\begin{itemize}}
\newcommand{\ei}{\end{itemize}}
\newcommand{\bea}{\begin{eqnarray}}
\newcommand{\eea}{\end{eqnarray}}
\newcommand{\f}{\frac}
\newcommand{\dg}{\dagger}
\newcommand{\la}{\langle}
\newcommand{\ra}{\rangle}
\newcommand{\ptl}{\partial}
\newcommand{\pr}{\prime}

\newcommand{\al}{\alpha}
\newcommand{\nn}{\nonumber}

\graphicspath{{../Figures/}}

\begin{document}
\title{Simulation of the Dynamics of Many-Body Quantum Spin Systems Using Phase-Space Techniques}
\author{Ray Ng}
\email{ngry@mcmaster.ca}

\author{Erik S. S{\o}rensen}
\affiliation{Department of Physics and Astronomy,
McMaster University
1280 Main Street West
L8S 4M1, Hamilton Ontario
CANADA}

\author{Piotr Deuar}
\affiliation{Institute of Physics, Polish Academy of Sciences (PAN),
Al. Lotników 32/46
02-668 Warszawa
Poland}

\date{\today}

\begin{abstract}
We reformulate the full quantum dynamics of spin systems using a phase space
representation based on SU(2) coherent states which generates an exact mapping
of the dynamics of any spin system onto a set of stochastic differential
equations. The new representation is superior in practice to an earlier phase space approach based
on Schwinger bosons, with the numerical effort scaling only linearly with system size. By also
implementing extrapolation techniques from quasiclassical equations to the full quantum
limit, we are able to extend useful simulation times several fold. This approach is
applicable in any dimension including cases where frustration is present in the spin system.
The method is demonstrated by simulating quenches in the transverse field Ising model in one
and two dimensions.  \end{abstract}

\maketitle

\section{Introduction}
\label{sec:intro}
With the development of cold atom experiments 
the non-equilibrium dynamics of closed quantum systems has become a focus of attention \cite{Polkovnikov:2011}.
In these experiments it has become feasible to prepare a model system in a specific eigen-state of $H_i$ and study the ensuing real-time
dynamics when the system evolves under a controllable Hamiltonian, $H_f$. This can be viewed as a realization of a quantum 
quench \cite{
CC1,CC2,Kollath07,Manmana07,Roux:2009,Roux:2010,Bernier:2011,
Trotzky:2012}.

Here we focus on how these effects occur in closed quantum spin systems \cite{Gobert:2005,Barmettler:2009,Barmettler:2010,Rossini:2009,Langer:2009,Calabrese:2011,Divakaran:2011,Essler12-I,Essler12-II,Essler:2012,Mitra:2013} neglecting couplings to the environment.
The dynamics of quantum spin systems is of particular interest for two reasons. First, they form a corner stone of condensed matter physics with many
open problems,  in particular for models with frustration, where even the equilibrium state is a matter of debate and little is known about the dynamics.
Secondly, using cold atom systems it has become conceivable to implement quantum simulators \cite{Feynman:1982,Porras:2004,Jaksch:2005,Lewenstein:2007,Buluta:2009,Johanning:2009,Schneider:2012}
using atomic degrees of freedom to mimic the quantum spin and their interactions.
Recent experiments \cite{Friedenauer:2008,Kim:2010,Simon:2011,Ma:2011,Islam:2011,Struck:2011,Kim:2011,Roos:2012,Britton:2012,Blatt:2012}
have shown significant progress towards realizing such a quantum simulator capable of simulating quantum spin systems.
Following the initial proposal \cite{Porras:2004} to implement such a simulator using trapped ions, it was experimentally	
realized with 2 spins \cite{Friedenauer:2008}, 3 spins \cite{Kim:2010} and up to 9 spins. \cite{Islam:2011,Kim:2011}
Recently, a system of $\sim$ 300 spins with Ising interactions were realized with trapped ions \cite{Britton:2012} and similar
system sizes have been reached using neutral atoms in optical lattices \cite{Struck:2011,Meinert:2013}.
As a model system, several of these experiments \cite{Friedenauer:2008,Simon:2011,Islam:2011,Britton:2012} model 
the transverse field Ising model (TFIM): \cite{TFIM}
\be
\hat{H}  = - J\sum_{\la i,j \ra} \hat{S}^{z}_{i}\hat{S}^{z}_{j}  - h(t) \sum_{i} \hat{S}^x_{i},
\label{eq:tfim}
\ee
which is the model that we focus on here.

Calculating the quantum dynamics of condensed matter spin systems is a
notoriously difficult problem, due to the macroscopic number of degrees of freedom. 
In this limit, the size of the Hilbert space scales exponentially deeming it
intractable in most cases. While some models can be solved analytically, they
are often not generalizable to higher dimension and the solution is often model specific.
For instance,  the TFIM \cite{TFIM} can be solved  exactly in one dimension using the Jordan-Wigner transformation
but this is not possible in higher dimensions. 

From a numerical perspective, the standard condensed matter computational
toolbox is remarkably successful but not completely general. For instance,
the direct `brute force' approach by way of exact diagonalization (ED), while always applicable, can
only accommodate relatively small system sizes of $N\sim 48$ (for a spin-$1/2$ system).
Quantum Monte Carlo (QMC)
methods \cite{Sandvik02} are extremely useful for calculating ground state
properties but only in the absence of any frustration. However, for the study of dynamics, QMC techniques are usually limited to imaginary times or
equivalently, imaginary frequencies.
Other methods, such as those rooted in the Density Matrix
Renormalization Group (DMRG) \cite{White:1992,DMRG1,DMRG2}, are the dominant techniques for one dimensional
systems \cite{White:1992} but are much harder to apply in two dimensions due to
scaling issues associated with the area law \cite{AreaLaw}. Currently, DMRG techniques are restricted to
one-dimensional systems and quasi two-dimensional strips.
Nonetheless, time dependent DMRG (tDMRG) \cite{tDMRG1,tDMRG2} has been very successful for one-dimensional systems
where the real-time dynamics of quantum spin systems can be treated out to $tJ/\hbar\sim 100$ \cite{Gobert:2005}.
Using time-evolving block-decimation (TEBD) \cite{TEBD}, the infinite size TEBD (iTEBD) \cite{iTEBD} has
yielded results out to $tJ/\hbar\sim 6-10$ \cite{Banuls:2009} for the TFIM and $tJ/\hbar\sim 20$ \cite{Barmettler:2009,Barmettler:2010}
for the XXZ spin chain and related models. It would therefore be quite worthwhile to explore techniques for calculating real-time dynamics
that are generally applicable to quantum spin systems in any dimension even in the presence of frustration. 

Another branch of numerical techniques fall under the category of quantum phase
space methods \cite{QuantumNoise, Corney04, DeuarPhD, Corney06, Drummond04, Polkovnikov10}.
They can be summarized by the following expression for the density operator
\begin{equation}
\label{eq: general density expansion}
\hat{\rho} = \int P(\vec{\lambda})  \hat{\Lambda} (\vec{\lambda})  d \vec{\lambda},
\end{equation}
where $\vec{\lambda}$ are parameters, $P(\vec{\lambda})$ plays the role of a
distribution and $\hat{\Lambda}(\vec{\lambda})$ is the operator kernel. Quantum
phase space methods have recently begun to gain exposure in condensed matter
systems. For instance, Polkovnikov\etal \cite{Polkovnikov10} have applied the path integral
formalism of the truncated Wigner representation to simulate quantum
quenches. Aimi\etal \cite{Aimi:2007a} extended the work by Corney\etal \cite{Corney04} by
successfully calculating the imaginary time dynamics of  the Hubbard
Hamiltonian in the high interaction limit using Fermionic Gaussian phase space
methods \cite{Aimi:2007b}. This was done by implementing symmetry projection
techniques \cite{Assaad:2005} as well as Monte Carlo methods. The high interaction limit was previously unattainable by QMC techniques.  

We focus specifically on the positive-P representation (PPR) which was developed by Drummond and co-workers \cite{Drummond80,Deuar02,Deuar06} 
and originally tailored to solve problems in quantum optics, where it has been applied with considerable success, as well as in ultracold bosonic gases.
For instance, Deuar\etal \cite{Deuar07}  have successfully implemented the PPR for the purpose of simulating multidimensional
Bose gases \cite{Deuar07,Deuar09,Deuar11}. 

The general idea is that the PPR provides an \textit{exact} mapping of the quantum
dynamics onto a set of Langevin type differential equations as long as boundary
terms do not arise \cite{Gilchrist97,Deuar02}. This mapping is made possible by the existence of
correspondence relations, which are characteristic of different phase space
methods. In principle, the PPR can be applied to both real and imaginary
times \cite{Drummond04} and since the computational effort is proportional to the system size it is
possible to simulate macroscopically large systems. In
addition, it is also possible to simulate frustrated systems which makes the PPR
particularly appealing. Finally, the PPR can be readily applied in any dimension.
The main drawback of the PPR however, is the possible
appearance of short simulation lifetimes signaled by the onset
of a divergences in the stochastic averages. 
Modified formalisms of the PPR based on the gauge-P
representation \cite{Deuar02,Deuar06b,Aimi:2007b,Deuar09b} have proven useful in this respect by allowing one to
introduce gauge functions that systematically remove unstable terms
in the stochastic differential equations (SDEs), and with it the source of divergences. This is typically
done at the expense of introducing an extra degree of freedom that plays the
role of a complex weight, $\Omega$.

The PPR formalism was first applied (to our knowledge) to the dynamics of many-spin
systems in our earlier work \cite{Ng11}, by treating the equivalent Schwinger
boson representation of spin chains with the canonical PPR method. 
It was
applied to the real time dynamics of one dimensional spin chains under a quantum quench.
A conclusion of that work was that the coherent-state basis used in
other studies where the PPR has been  successful is not very suitable for systems
composed of $S=1/2$ quantum spins. It led to both early noise onset, and the need
for a broad initial distribution to describe the number state that corresponds
to $S=1/2$ spin.  The latter issue is particularly onerous as it turns out
to preclude efficient sampling of the distribution for large numbers of
spins ($\mc{O}(100)$) --- the regime where phase-space methods are particularly advantageous.

In this paper, we choose a different route and
describe the system using the SU(2) basis \cite{Radcliffe71,Zhang:1990,Barry08}. This allows us to 
develop a PPR-like distribution, which is then used to obtain stochastic
differential equations that do not suffer from the broad distribution and sampling issues
encountered with PPR in Schwinger Bosons. 
A related approach has been used in the past on an imaginary time evolution of the Ising model by Barry\etal \cite{Barry08} 
using an unnormalized kernel for the density operator. We will however use a normalized kernel, which is  more appropriate for simulating dynamics \cite{DeuarPhD}. 

The outline of the paper is as follows.
The SU(2) coherent state phase-space representation and related formalism is derived in Sec.~\ref{sn: formalism}.
Its basic application to the TFIM is discussed in Sec.~\ref{sn: bare TIM}.
Even though this new approach leads to significantly longer simulations times,
limitations are clearly present and we also discuss these later in the section. 
It is possible to extend the simulation time even further by extrapolating from regimes
with reduced quantum fluctuations into the full quantum regime. This \emph{entanglement scaling} technique
is described in Sec.~\ref{sn: extrapolation}. This allows us to obtain longer simulation times, similar to typical timescales of the problem. We
then conclude in section~\ref{sn: conclusions} and discuss the
future direction of this work.
Some more technical aspects are relegated to appendices.

\section{The formalism}
\label{sn: formalism}
\subsection{The SU(2) basis}
\label{sn: first formalism}
Traditionally, the PPR formalism is based upon bosonic coherent
states \cite{Glauber63}, and hence the most natural generalization to spin
systems would be the use of SU(2) coherent
states \cite{Radcliffe71,Arecchi72,Perelomov72,Zhang:1990}. The bosonic coherent states and SU(2) coherent states
are analogous and
have similar properties such as that of overcompleteness and in the large $S$ limit the SU(2) coherent states approach
the bosonic coherent states. 
While the spin versions of other kinds of
  phase-space representations (the Q
      representation \cite{Shastry78,Lee84,Drummond84}, P
      representation \cite{Narducci74}, and Wigner
      representation \cite{Chaturvedi06}) have been introduced in the past, the
  advantage of the PPR approach is that the kernel can be made analytic in the
  phase-space variables, which in turn guarantees that standard stochastic
  diffusion equations can be obtained for the evolution \cite{Drummond80}. 

Labeling the spin quantization direction as $\vec{z}$, with operator $\hat{S}^z$, we define the SU(2) coherent states for a spin $S$\cite{Radcliffe71,Arecchi72,Perelomov72,Zhang:1990}:
\be
\label{eq: SU2 coherent state}
| z \ra = e^{-zS}e^{e^{z} \hat{S}^{\dagger} } | S, -S \ra,
\ee
where $S^{\dg}$ is the raising operator and $|S,S_z \ra$ is the state with $S_z$ spin projection onto the quantization direction. 
The state is parametrized by a single complex variable $z$ (not to be confused with the quantization direction $\vec{z}$).
Our interest lies in the spin-$\f{1}{2}$ case for which (\ref{eq: SU2 coherent state}) reduces to the SU(2) case:
\be
\label{eq: spin-1/2 coh state}
| z \ra = e^{-\f{z}{2}}e^{e^{z} \hat{S}^{+}} | \downarrow\ra  = \left[\ba{c}  e^{z/2} \\ e^{-z/2} \ea \right].
\ee
with $|\!\downarrow\ra=\left[\ba{c}0\\1\ea\right]$, and $S^+=\left[\ba{cc}0&1\\0&0\ea\right]$.
In this case,  $|z\ra$ has the physical interpretation of being a unit vector pointing to the  position $(\theta, \phi)$ on the surface of the Bloch sphere. 
The transformation that relates the $z$-coordinate to $(\theta, \phi)$ is  $e^z = e^{i \phi} \tan( \theta/2) $ where $\theta\in [0,\pi/2]$ is the polar angle and $\phi\in[0,2\pi]$ is the azimuthal one. 


\subsection{SU(2) phase-space representation}
\label{ssn: representation}
To obtain stochastic evolution equations with positive diffusion, we follow standard PPR procedure \cite{Drummond80, QuantumNoise, Drummond05}. For brevity, we will only highlight key aspects of the formalism and refer interested readers to \cite{Ng11} where a spin system is worked out in detail. 

First, we represent the density matrix (\ref{eq: general density expansion}) using an off-diagonal kernel with unit trace constructed from SU(2) coherent states:
\be
\label{eq: kernel}
\hat{\Lambda}  = \f{| z \ra \la z^{\pr \ast} | } {  \la z^{\pr\ast } |   z \ra   }.
\ee
For an $N$-site system, one uses a tensor product of independent kernels for each site $i$:
\be
\label{eq: kernelN}
\hat{\Lambda}  = \bigotimes_{i=1}^{N}\ \hat{\Lambda}_i(z_i,z^{\pr \ast}_i).
\ee
 It is parametrized by the set $\vec{\lambda} = \{z_1,\dots,z_N,z'_1,\dots,z'_N\}$ of $2N$ independent complex variables.

The dynamics of any system is obtained by evolving the equation of motion for the density operator,
\be
\label{eq: master eqn}
\f{d\hat{\rho}}{ dt} = -\f{i}{\hbar} \left[ \hat{H}, \hat{\rho} \right].
\ee
To counteract the exponential complexity of the Hilbert space with large system sizes, there exists an equivalent description of (\ref{eq: master eqn}) in terms of a Fokker-Planck equation (FPE) for the distribution function $P(\vec{\lambda})$ (c.f. (\ref{eq: general density expansion})) in the continuous space of the phase-space variables $\vec{\lambda}$. This in turn can be mapped onto stochastic equations for the variables which is the final result of the formalism. These can be sampled with a chosen ensemble whose size $\mc{N}$ controls the numerical effort, trading it off for statistical precision. 

The FPE is obtained by using correspondence relations that establish a duality between the action of spin operators and differential operators on the kernel, $\hat{\Lambda}$.  It is possible to show that spin operators acting from the left of the kernel satisfy the following identities (site index $i$ implied):
\bea
\label{eq: left cr sx}
\hat{S}^{x}\hat{\Lambda} & =&  \left[ -\sinh z \f{\ptl}{\ptl z} + S^x  \right] \hat{\Lambda}  \\
\label{eq: left cr sy}
\hat{S}^{y}\hat{\Lambda} & =&  \left[ -i\cosh z \f{\ptl}{\ptl z} + S^y \right] \hat{\Lambda} \\
\label{eq: left cr sz}
\hat{S}^{z}\hat{\Lambda} & =&  \left[\f{\ptl}{\ptl z} + S^z  \right]\hat{\Lambda} 
\eea
while spin operators acting from the right satisfy
\bea
\label{eq: right cr sx}
\hat{\Lambda}\hat{S}^{x} & =&  \left[ -\sinh z^\pr \f{\ptl}{\ptl z^\pr} + S^{ x\pr} \right] \hat{\Lambda}    \\
\label{eq: right cr sy}
\hat{\Lambda}\hat{S}^{y} & =&  \left[ i\cosh z^\pr \f{\ptl}{\ptl z^\pr} + S^{ y \pr} \right] \hat{\Lambda}    \\
\label{eq: right cr sz}
\hat{\Lambda}\hat{S}^{z} & =&  \left[\f{\ptl}{\ptl z^\pr} + S^z  \right]\hat{\Lambda},
\eea
where 
\bea
\label{eq: Sxest}
S^{x} & = & \  \f{1}{2} ( \cosh z - \sinh z \tanh R  ) \\
\label{eq: Syest}
S^{ y} & = & \  \f{i}{2} \left( \sinh z - \cosh z \tanh R  \right) \\
\label{eq: Szest}
{S}^z & = & \f{1}{2} \tanh(R).
\eea
and
\be
 R = (z + z^\pr)/2.
\ee
The primed counterparts of (\ref{eq: Sxest})-(\ref{eq: Syest}) are easily obtained by making the substitutions $z\rightarrow z^\pr, i \rightarrow -i$, so that $(S^{y})^*=S^{y\pr}$ when $z=z^{\pr\ast}$.

To derive estimators for expectation values $\la\hat{O}\ra$ of general observables $\hat{O}$, we start from the usual expression:
\be
\label{eq: observable}
\la\hat{O}\ra = \f{{\rm Tr}[\hat{O} \hat{\rho} ]}{{\rm Tr}[\hat{\rho}]}   = \f{\la\la {\rm Tr}[\hat{O}\hat{\Lambda}]\ra\ra}{\la\la{\rm Tr}[\hat{\Lambda}]\ra\ra},
\ee
where the right term follows from (\ref{eq: general density expansion}), with $\la\la\cdots\ra\ra$ denoting an average over the ensemble that samples $P(\vec{\lambda})$, i.e. $\la\la  \dots \ra\ra  =  \int P(\hat{\lambda}) (\dots) d\vec{\lambda}$. Noting that ${\rm Tr}[\hat{\Lambda}] = 1$ and using (\ref{eq: left cr sx})-(\ref{eq: left cr sz}), one obtains e.g.
\be
\label{eq: obsS}
\la\hat{S}^\al\ra  = \la\la S^\al\ra\ra, 
\ee
for $\al=x,y,z$. This explains the choice of notation $S^{x,y,z}$ in (\ref{eq: left cr sx})-(\ref{eq: Szest}). Using the cyclic property of the trace in (\ref{eq: observable}) and (\ref{eq: right cr sx})-(\ref{eq: right cr sz}), one could have just as well have derived the equivalent estimators for the spin components using primed coordinates instead:  $\la\hat{S}^{\al }\ra = \la\la S^{\al \pr} \ra\ra$. Either estimator is valid but in our calculations we chose to use (\ref{eq: obsS}) simply as a matter of preference. Notably, since the kernel is normalized the expectation value of its derivative is zero and  one can obtain estimators for more complex observables  by taking the expectation value of appropriate correspondence relations.  Once the FPE is obtained, its mapping onto Ito SDEs is well known \cite{QuantumNoise} and in doing so, we effectively map the dynamics of $N$-spins onto $\sim N$ complex variables: $\{ \vec{\lambda} \}$.

An ensemble of $\mathcal{N}$ realizations $\{ \vec{\lambda}^{(i)} \}$, with $i=1,\dots,\mathcal{N}$ becomes equivalent to the full quantum mechanical description of the system as the ensemble size becomes large ($\mathcal{N}\to\infty$). In practice, $10^3 - 10^6$ trajectories are typically sufficient for good convergence, depending on the desired precision.




\subsection{Stochastic equations for quantum dynamics}
\label{ssn: quantum dynamics}



Even though the numerical results that we present later are only for the transverse field Ising model, (\ref{eq:tfim}), it is instructive to consider the stochastic equations for slightly
more general models. For generality, we therefore consider the Heisenberg Hamiltonian in a transverse field $h(t)$ along the $\vec{x}$ direction:
\be
\label{eq: Hamiltonian-general}
\hat{H}  = - J\sum_{\la i,j \ra} \left[ \hat{S}^{z}_{i}\hat{S}^{z}_{j}  + \Delta\left( \hat{S}_{i}^{y}\hat{S}_{j}^{y} + \hat{S}_{i}^{x}\hat{S}_{j}^{x}\right) \right] - h(t) \sum_{i} \hat{S}^x_{i},
\ee
with each connected-neighbor pair $\la i,j\ra$ counted once. Here $J$ is the hopping strength ($J>0$  for the ferromagnetic system), $\Delta$ governs in-plane anisotropy,  $h(t)$ is the transverse field strength, and we choose units such that $\hbar=1$.
Following section~\ref{ssn: representation}, we derive Ito stochastic equations to describe the dynamics of the system:
\begin{widetext}
\bea
\f{dz_{i}}{dt} &  = &    \f{iJ}{2}  \sum_{j\in{\rm n}(i)}\tanh R_j- i h(t) \sinh z_{i} + \sqrt{J}\left[\sum_{j\in{\rm n_L}(i)}\eta_{\la i,j\ra} + i\sum_{j\in{\rm n_R}(i)} \eta^*_{\la i,j\ra}\right] \nn\\
&& \qquad - i\f{\Delta J}{2}  \sum_{j\in{\rm n}(i)}\left(\mc{S}_{ij}+\mc{C}_{ij}\tanh R_j\right) + \sqrt{J\Delta}\left[\sum_{j\in{\rm n_L}(i)} \sqrt{\mc{C}_{ij}}\ \xi_{\la i,j\ra} 
-i \sum_{j\in{\rm n_R}(i)} \sqrt{\mc{C}_{ij}}\ \xi^*_{\la i,j\ra}\right]  \label{dzdt}\\
\f{dz^{\pr}_{i}}{dt} &  = &    -\f{iJ}{2}  \sum_{j\in{\rm n}(i)}\tanh R_j+ i h(t) \sinh z^{\pr}_{i} + \sqrt{J}\left[\sum_{j\in{\rm n_L}(i)}\eta^{\pr}_{\la i,j\ra} -i \sum_{j\in{\rm n_R}(i)} \eta^{\pr*
}_{\la i,j\ra}\right] \nn\\
&& \qquad + i\f{\Delta J}{2}  \sum_{j\in{\rm n}(i)}\left(\mc{S}^{\pr}_{ij}+\mc{C}^{\pr}_{ij}\tanh R_j\right) + \sqrt{J\Delta}\left[\sum_{j\in{\rm n_L}(i)}  \sqrt{\mc{C}^{\pr}_{ij}}\ \xi^{\pr}_{\la i,j\ra} 
+i \sum_{j\in{\rm n_R}(i)} \sqrt{\mc{C}^{\pr}_{ij}}\ \xi^{\pr*}_{\la i,j\ra}\right]  \label{dzidt}
\eea
\end{widetext}
The $R,\mc{C}$, and $\mc{S}$ functions are
\be
R_{i} = \frac{z_i+z'_i}{2}
\ee
\bea
{\mc C}_{ij} = \cosh(z_i- z_j) &\qquad& {\mc S}_{ij} = \sinh(z_i-z_j)\\
{\mc C}^{\pr}_{ij} = \cosh(z^{\pr}_i- z^{\pr}_j) &\qquad& {\mc S}^{\pr}_{ij} = \sinh(z^{\pr}_i-z^{\pr}_j)
\eea
The noise $\eta,\xi,\eta',\xi'$ takes the form of complex Wiener increments  of zero mean, one of each per connected pair $\la i,j\ra$. 
They are all independent of each other, and delta-time-correlated. That is, the only nonzero second order moments are
\be\label{noisevars}
\la\la x_{\la i,j\ra}(t) x_{\la i,j\ra}^*(t')  \ra\ra  = \delta(t- t^{\pr})
\ee
where $x$ can stand for any symbol in $\left\{ \eta,\xi,\eta',\xi' \right\}$. 
Individual complex noises are easily constructed in practice from two real Gaussian random variables of variance  $\frac{1}{2\Delta t}$  at each time step of length $\Delta t$ (one for the real, one for the imaginary part).

Some notation is also required to keep track of the connectivity: ${\rm n}(i)$ indicates the set of connected neighbors for site $i$. For example, a 1D chain with nearest-neighbor coupling has ${\rm n}(i) = \{i-1,i+1\}$. 
The noises couple connected sites in such a way that when one member of the pair gets the complex noise $\eta$, the other gets $i\eta^*$ or $-i\eta^*$ depending on the details. 
Hence, if we assign to each such bond $\la i,j\ra$ an arbitrary labeling directionality $i\to j$, then the ``left'' site $i$ gets $\eta_{\la i,j\ra}$ noise while the ``right'' site $j$ gets the conjugate one. The neighbors that are labeled as ``left'' sites for the $\la i,j\ra$ bond are in the set ${\rm n_L}(i)$, while those that are labeled as ``right'' sites are in the  set ${\rm n_R}(i)$. For the 1D example, one can have ${\rm n_L}(i) = \{i-1\}$ and ${\rm n_R}(i) = \{i+1\}$.
With this notation, the expressions (\ref{dzdt})-(\ref{dzidt}) allow for arbitrary connectivity between the sites, including frustrated systems.


The equations (\ref{dzdt})-(\ref{dzidt}) are equivalent to the Schwinger boson phase-space stochastic equations developed in \cite{Ng11}, but their statistical properties at finite but large ensemble size $\mc{N}$ are very different. Importantly, in the present representation any product state $\otimes_i |z^0_i\ra$ can be described as a delta function distribution 
\be\label{ic}
P(\vec{\lambda}) = \prod_i \delta^{(2)}(z_i-z^0_i) \delta^{(2)}(z_i-z^{\pr *}_i).
\ee
This can be used to initialize the $t=0$ ensemble in a simple fashion. 
More importantly, since this is a zero-width distribution, the initial state remains well sampled and compact even for very large systems. 

A technical hurdle is encountered for the exact $|\!\uparrow\ra$ and
$|\!\downarrow\ra$ states, which correspond to the limit $z\to\pm\infty$,
  respectively. Some cut is always required when mapping the
  surface of a sphere (such as the Bloch sphere for the spin-$\f{1}{2}$ case)
  onto a plane, and the in-plane evolution at a cut is singular. In our $z$ mapping there are two cuts like in a cylindrical map projection. 
We deal with this issue by re-projection onto a
polar coordinate plus a Boolean variable that keeps track of which pole
is being used to define the coordinates.  
This is explained in Appendix~\ref{sn: new vars}.


\subsection{Thermal calculations}
\label{ssn:thermal}
An imaginary time evolution in the temperature variable $\beta=1/k_BT$ can also be formulated in principle using the anticommutator \cite{Drummond04}: $d\hat{\rho}/d\beta = -\frac{1}{2}\left[\hat{H}\hat{\rho}+\hat{\rho}\hat{H}\right]$. 
For the simple ferromagnetic 1D Ising model ($J=1, h=0$) considered previously in this context \cite{Barry08}, we obtain, for comparison: 
\bea
\f{dz_{i}}{d(\beta/2)}  & = &   \f{1}{2}  \left[  \tanh R_{i-1} + \tanh R_{i+1}  \right]  +  \eta_{i} + \eta_{i-1}^{\ast} \quad\\
\f{dz_{i}^{\pr}}{d(\beta/2)} &  = &   \f{1}{2}  \left[  \tanh R_{i-1} + \tanh R_{i+1} \right]    + \eta^{\pr}_{i} + \eta_{i-1}^{\pr \ast} \quad\\
\f{dW}{d(\beta/2)} &=& \f{1}{2}W\sum_i \left\{ \tanh R_i \tanh R_{i+1}\right\}.
\eea
with noise variances $\la\la\eta_i(\beta)\eta_j^*(\beta')\ra\ra=\delta(\beta-\beta')\delta_{ij}$.  
The variable $W$ is a trajectory-dependent weight and has to be taken into consideration in Eq.~(\ref{eq: observable}). For a general observable $\hat{O}$, $\la \hat{O}\ra$ is  now given by 
\be
\label{eq: wt obs}
\la \hat{O} \ra = \f{ \la \la W O(\vec{\lambda}) \ra \ra}{\la \la W \ra\ra},
\ee
where $O(\vec{\lambda})$ represents the stochastic estimator that is a function of phase space variables $\vec{\lambda}$. 
Note that the energy units we choose here are a factor of two smaller than in \cite{Barry08}, so that $\beta/2$ is the imaginary time used there. In comparison, the noise terms are the same, but the normalized kernel we use introduces the $\tanh R$ drift terms and evolving weights.

A $T=\infty$ initial condition $\hat{\rho}=I/N$ can be obtained in a number of ways. One can have e.g. a uniform distribution of $z_i$ on the imaginary axis on $[-\pi,\pi]$ as in \cite{Barry08}, or an even random mix of $z=\pm z_0$ with $z_0\to\infty$.  In both cases, $z'_i=z_i^*$.  Such freedom is typical for overcomplete representations, and may lead to different statistical properties depending on the initial distribution chosen. 

\section{Demonstration of the basic method}
\label{sn: bare TIM}

\subsection{Transverse field quench}
\label{ssn: quench results}
We now apply our formalism to the dynamics of a transverse field quench of the ferromagnetic Ising model ($J=1, \Delta=0$), Eq.~(\ref{eq:tfim}).
This model is a realistic description of many physical phenomena \cite{TFIM}, and with recent advances in ultra cold atoms and the high degree of parameter control it is now possible to reproduce quenches
in isolated quantum systems and to study the ensuing unitary dynamics. 
In this context the TFIM is of considerable interest as a model system.
There has been much recent work done on this system both theoretically  
\cite{Rossini:2009,Essler12-I, Essler12-II}
and experimentally \cite{Friedenauer:2008,Coldea:2010,Simon:2011,Islam:2011,Britton:2012}. 


The quench occurs at $t=0$ with a time-dependent field given by
\bea
h(t) = \left\{ \ba{c}  0 ,\ \  t\leq 0 \\ \\ h,\ \  t > 0\ea \right. 
\eea
We choose to start from the $h=0$ spin-up ground state$|\!\uparrow \uparrow \dots \uparrow \ra$ and 
quench a 1D spin chain to a value of 
\be
h=h_c=0.5. 
\ee
This is the well known critical point of the spin model, where the correlation length in equilibrium diverges\cite{Pfeuty79}, separating the ferromagnetic and paramagnetic phases.

\begin{figure}
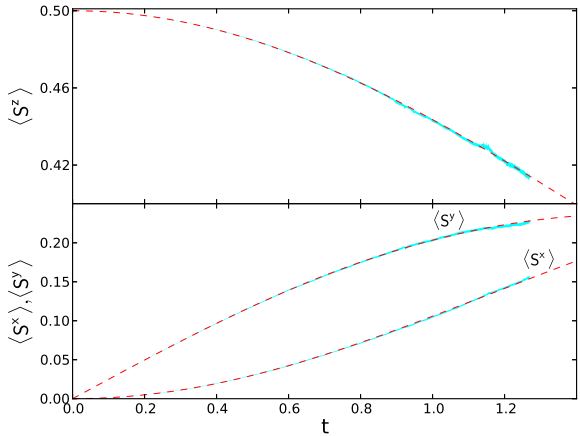

\centering
\includegraphics[width=\columnwidth]{{{figure1}}}
\vspace*{-0.3cm}
\caption{(color online).  Spin components: $\la\hat{S}^x \ra, \la\hat{S}^y \ra , \la\hat{S}^z \ra$ vs. time 
 for the ten-site $1D$ Ising spin chain with transverse quench from $h=0$ to $h_c=0.5$. 
Red dashed lines show exact diagonalization results. Our calculations including  error bars are indicated by the cyan region. 
\label{fig:1}}\vspace*{-0.3cm}
\end{figure}

\begin{figure}
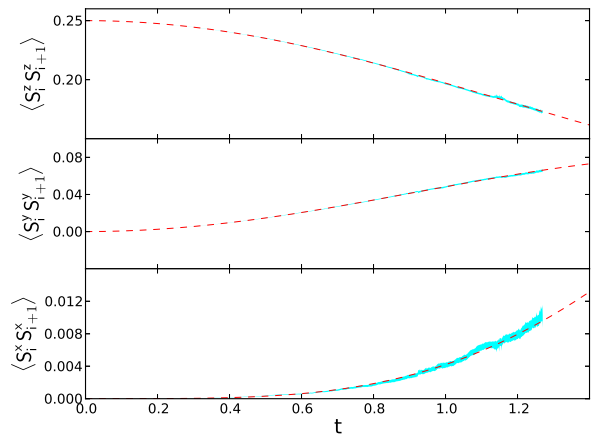

\centering
\includegraphics[width=\columnwidth]{{{figure2}}}\vspace*{-0.3cm}
\caption{(color online).  Nearest neighbor correlation functions 
$\la\hat{S}^x_i\hat{S}^x_{i+1} \ra, \la\hat{S}^y_i\hat{S}^y_{i+1} \ra , \la\hat{S}^z_i\hat{S}^z_{i+1}  \ra$ 
vs. time  for a 1D ten-site Ising spin chain with transverse quench from $h=0$ to $h_c=0.5$. 
Red dashed lines show exact diagonalization results. Our calculations including  error bars are indicated by the cyan region. 
\label{fig:2}}\vspace*{-0.3cm}
\end{figure}

Rewriting Eqs.~(\ref{dzdt})-(\ref{dzidt}) we find that for the 1D TFIM the equations to simulate are 
\bea
\f{dz_{i}}{dt} &  = &   \ \ i \left[S^z_{i-1} + S^z_{i+1}\right] - i  h\sinh z_{i} + \eta_i + i\eta^*_{i-1}\label{dzdt1d}\\
\f{dz^{\pr}_{i}}{dt} &  = &    -i \left[S^z_{i-1} + S^z_{i+1}\right] + ih  \sinh z^{\pr}_{i} + \eta^{\pr}_i -i\eta^{\pr*}_{i-1}\label{dzidt1d}
\eea
with estimators (\ref{eq: Szest}).  We calculate the dynamics of the  expectation values of spins $\la\hat{S}^\al\ra$, and nearest-neighbor spin correlations $\la \hat{S}^\al_i \hat{S}^\al_{i+1} \ra  $ in the three orthogonal axis directions: $\al=\{x,y,z \}$.
Our initial results are shown  in Figs.~\ref{fig:1} and \ref{fig:2} for an $N=10$ site chain that is small enough that exact results by way of diagonalization are available for comparison. The stochastic averages are in excellent agreement with the exact results. 
We use $\mc{N}=10^4$ trajectories distributed among $\mc{B}=100$ equal sized bins. 
The statistical uncertainty in the estimators for observables can then be determined with the help of the central limit theorem, i.e. the  errorbars in the final estimates  are obtained by averaging over all bins is $1/\sqrt{B}$ times the standard deviation of the $\mc{B}$ single-bin-averaged estimators.  

We observe the onset of spiking after a certain time, $t_{\rm sim}$, which is a known feature of some PPR-like calculations when the equations are nonlinear. This is also often a sign of the onset of sampling difficulties \cite{Gilchrist97}. 
Simulations are stopped at $t_{sim}$ which we determine by 
the criterion (\ref{tanshing}) (see Appendix~\ref{sn:spikes} for details).
This time compares favorably to the simulation time of $t_{\rm sim}\approx0.6$ seen in our earlier Schwinger boson calculations \cite{Ng11}. 

\begin{figure}
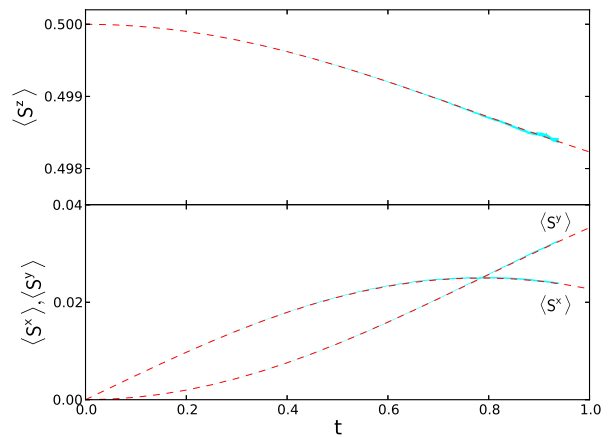

\centering
\includegraphics[width=\columnwidth]{{{figure3}}}
\vspace*{-0.3cm}
\caption{(color online).  Spin components: $\la\hat{S}^x \ra, \la\hat{S}^y \ra , \la\hat{S}^z \ra$ vs. time 
 for the 2D Ising spin model on a $3\times3$ square lattice with transverse quench from $h=0$ to $h=0.1$.  Red dashed lines show exact diagonalization results. }
\label{fig:3}\vspace*{-0.3cm}
\end{figure}

Fig.~\ref{fig:3} shows a calculation for a 2D system on a $3\times3$ square lattice. Again we use a small system size of $3\times3$ to allow comparison with exact diagonalization. 
Much larger systems can be treated, as will be demonstrated in Fig.~\ref{fig:others} in Section~\ref{sn: extrapolation}. 


\subsection{Limitations on simulation time}
\label{ssn:limitations}
While real time simulations now last longer than in \cite{Ng11} and scale well with system size even in higher dimensions (see Figs.~\ref{fig:others}), it would be very desirable to obtain much longer simulation times. 

A major stumbling block is that at the points in phase-space where $R_i = \pm i\pi/2$, the factor $\tanh(R_i)=2S^z_i$ diverges. This is a problem as it appears both in observable calculations (\ref{eq: Sxest})-(\ref{eq: Szest}) and in the evolution equations (\ref{dzdt1d})-(\ref{dzidt1d}). In observables, spiking  appears when a trajectory passes close to a pole, which obscures the mean result when it happens often. In the evolution equations, this causes a poorly integrated sudden jump, and in fact can be a symptom of the onset of systematic errors \cite{Gilchrist97}. In the present case, these poles are at the root of the limitations on simulation time. 

It is helpful to look at the equations for $R$ and a complementary independent variable
\be
Q_i = \f{z_i-z_i^{\pr}}{2i}.
\ee
 Consider for now what happens if the transverse field $h$ is turned completely off, the equations are

\bea
\f{dR_i}{dt} &=& \frac{1}{2}\left[ \eta_i + i\eta_{i-1}^* + \eta'_i -i\eta_{i-1}^{\pr *}\right]\\
\f{dQ_i}{dt} &=& \frac{1}{2}\left[  \tanh R_{i-1} + \tanh R_{i+1}  \right] \nn\\&&\qquad- \frac{i}{2}\left[ \eta_i + i\eta_{i-1}^* -\eta'_i + i\eta_{i-1}^{\pr *}\right].
\eea
The evolution of $Q_i$ becomes singular when either of the  $R_{i+1}\text{ or }R_{i-1}= \pm i\pi/2$. For a small deviation $\delta_{i+1}$ or $\delta_{i-1}$ from such a pole, as in e.g. $R_{i+1}=\pm i\pi/2 + \delta_{i+1}$, we have 
\be
\f{dQ_i}{dt} \approx \frac{1}{2\delta_{i+1}} + \text{noise}.
\ee
 The evolution of $R$, on the other hand, is purely complex diffusion, with variances ${\rm var}(|R|) =t$. Thus, even if there is no transverse field $h$, some trajectories will eventually diffuse from $z_0$ onto the $\pm i\pi/2$ poles in a time $\propto( z_0^2+\pi^2/4)$. 
We see that in the ground-state limit of $z_0\to\infty$, this time becomes ever longer. For finite $h$ values, there is also a more rapid deterministic drift away from the $h=0$ ground state due to precession induced by the transverse field. An analysis of $t_{\rm sim}$ that takes into account finite $h$ values is given in Appendix~\ref{sn:h}.

While a fundamental resolution or alleviation of these issues for spin states is beyond our scope here, there is a fairly straightforward procedure that one can use to extract physical information for appreciably longer times than those seen in Figures.~\ref{fig:1}-\ref{fig:hdep}. It 
is described and demonstrated in the following Section~\ref{sn: extrapolation}. 






\subsection{Simulation time in the SU(2) basis}
\label{ssn:tsimh}
\begin{figure}
\centering
\includegraphics[width=0.7\columnwidth]{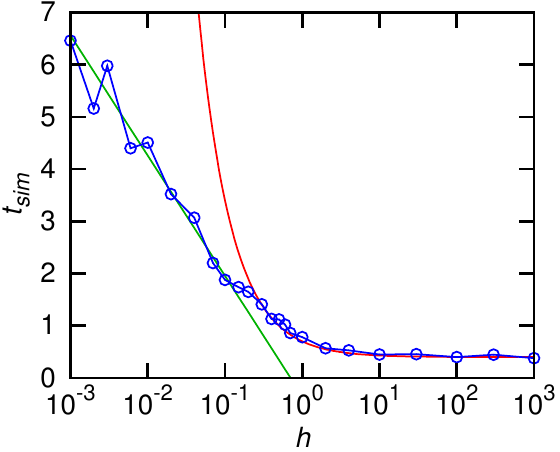}\vspace*{-0.3cm}
\caption{(color online).  
The dependence of simulation time $t_{\rm sim}$ on the quench strength $h$ (blue) for the 1D transverse quench Ising model ($\Delta=0$). The trends at low $h$, Eq.~(\ref{hdeplo}) (green line) and at large $h$, Eq.~(\ref{hdephi}) (red line) are also shown.  
\label{fig:hdep}\vspace*{-0.3cm}
}
\end{figure}

Figure~\ref{fig:hdep} shows the dependence of the simulation time $t_{\rm sim}$ on the quench strength $h$. The trends are logarithmic at small $h$,
\be\label{hdeplo}
t_{\rm sim} \sim \frac{2}{C}\log \frac{c_0\sqrt{C}}{h},
\ee
and approximately constant
\be\label{hdephi}
t_{\rm sim} \approx c_1\frac{1}{C} + \frac{c_2}{h}
\ee
for large $h$. $C$ is the number of connections per site ($C=2$ here), and the constants are $c_0\approx0.5$, $c_1\approx 0.8$ and $c_2\approx0.3$.
These trends are derived in Appendix~\ref{sn:h}. The simulation time in both regimes is inversely proportional to $C$, hence also to the dimensionality $d$.


\subsection{Origin of the poles}
\label{ssn: originpoles}
For future work in the field, it is instructive to understand why such poles appear in phase-space in the first place. 
Consider the matrix representation of $\hat{\Lambda}$: 
\be
\hat{\Lambda} = \frac{1}{2\cosh R}\left[\begin{array}{cc}
e^R & e^{iQ} \\ e^{-iQ} & e^{-R}
\end{array}\right].
\ee
Projectors onto pure states $|z\ra\la z|$ correspond to $z'=z^*$, and thus to real values of $R$ and $Q$. This is the set of all hermitian kernels, and all such kernels are well behaved. However, non-hermitian kernels that contain complex $R$ can be singular if the denominator in the normalization approaches zero. The worst case occurs when $\cosh R=\frac{1}{2}\la z|z'^*\ra = 0$, i.e. when the kernel causes a transition between orthogonal states. This is the exact location of the unwanted poles in the equations mentioned in the previous section, i.e. when $R_i = \pm i n^\pr\f{\pi}{2}$ and $n^\pr \in \text{odd}$.  Such unwelcome behavior occurs for these states because the present kernel $\hat{\Lambda}$ was explicitly normalized to have unit trace, while such coherences between orthogonal states have zero trace. They are un-normalizable, and pathological behavior ensues. 

We have also attempted the obvious idea to use an un-normalized kernel $|z\ra\la z^{\pr *}|$ which never divides by a zero trace. However, despite having equations of motion with no divergent terms ($\tanh R$ or otherwise), this representation produces a simulation in which systematic errors grow linearly right from $t=0$. 
The cause are ``Type-II'' boundary term errors of the kind described in \cite{DeuarPhD}: observable calculations (\ref{eq: observable}) now involve ensemble averages of 
complex weight factors ${\rm Tr}[\hat{\Lambda}] = W = 2\cosh R$ e.g. $\la\hat{S}_z\ra = \la\la W S^z\ra\ra / \la\la W \ra\ra$, as in (\ref{eq: wt obs}). 
The exponential nature of the weight factors leads them to be poorly sampled. This is  because the distribution of $W(z,z')$ has very different behavior than the actual Gaussian sample distribution that generates the noise in the evolution equations for $z$ and $z'$. In particular, trajectories with ${\rm Re}[R]$ several standard deviations above the mean are never generated, while their contribution to weights $W(R)$ included may be significant. 

The dynamical and normalization behavior described above bears resemblance to similar afflictions seen in PPR simulations of the bosonic anharmonic oscillator $\hat{H}=\hat{a}^{\dagger 2}\hat{a}^2$ \cite{DeuarPhD,Deuar06}. There, a variable $n$ whose real part is averaged to obtain the occupation number $\la\hat{a}^{\dag}\hat{a}\ra=\la\la n\ra\ra$ takes on complex values in the course of the evolution. Unstable regions of phase space are accessed through diffusion into the imaginary part of $n$, much as here diffusion into the imaginary part of $R$ sets off an instability. Similarly, an un-normalized kernel for the anharmonic oscillator alleviates instability, but makes observable calculations suffer again from Type-II boundary terms right from $t=0$. 

This, and past work on Bose systems treated with the original positive-P representation allows us to speculate that such effects are generic features of PPR-like phase-space methods with analytic kernels constructed from off-diagonal  basis states:
\begin{enumerate}
\item Complex parts of variables whose real parts correspond to physical observables mediate instability.
\item Systematic errors, or at least huge noise, tend to ensue when phase-space evolution accesses regions corresponding to kernels with zero trace.  
\item The use of an un-normalized kernel is not effective, as type-II boundary term errors in the observable calculations tend to result. 
\end{enumerate}

\section{Extended simulation time by entanglement scaling}
\label{sn: extrapolation}

\subsection{Entanglement scaling}
We will apply a technique developed for many-body simulations of Bose gases in tandem with the PPR \cite{Deuar09}, that uses the trend of results from calculations with reduced noise terms to pinpoint the full quantum values. 
Such a trend can be useful because reduced noise leads to longer simulation times before the onset of spiking.  
We will call this approach ``entanglement scaling'' because it is the noise that is responsible for generating new entanglement between the sites. Recall that the kernel is separable, so all entanglement in the system is described by the distribution. Noiseless equations produce no entanglement.

To use the technique, we need several families of stochastic simulations (labeled $m=A,B,\dots$), parametrized by variables $\lambda_m\in[0,1]$, that interpolate
 smoothly between long-lasting, reduced-noise equations at $\lambda_m=0$ and the full quantum description at $\lambda_m=1$. At least two independent families are required to assess the accuracy of trends extrapolated to $\lambda_m=1$. 
Technical details are summarized in Appendix~\ref{ssn: extrap-proc}. The philosophy of this approach is similar to comparing trends of results obtained with different summation techniques in diagrammatic Monte Carlo \cite{Prokofev08}.

The first family of equations, A, will be the SU(2) equations (\ref{dzdt1d})-(\ref{dzidt1d}) with noise terms multiplied  by $\sqrt{\lambda_A}$, so that $\lambda_A=0$ gives completely noiseless equations with no entanglement. 
Scaling noise variance linearly with $\lambda_A$ here tends to give observable estimates that are also nearly linear in $\lambda_A$. This aids in making the extrapolation of the trend to $\lambda_A=1$ well conditioned, since few fitting parameters are needed.

The second family, family B will use the same noise $\eta_i$ for both $z_i$ and $z_i^{\pr\ast}$ variables at $\lambda_B=0$. The difference in stochastic equations between $\lambda_B=0$ and $\lambda_B=1$ in this family is analogous to 
that between equations for a boson field under a Glauber-Sudarshan P representation \cite{Glauber63b,Sudarshan63} and a positive-P representation, respectively. At $\lambda_B=0$ one now has stable, albeit stochastic equations, but they do not correspond to full quantum mechanics. 
The following choice of $\lambda_B$-dependence gives approximately linear scaling of observable estimates with $\lambda_B$:
\be
\eta'_i = \sqrt{\lambda_B(2-\lambda_B)}\,\widetilde{\eta}_i + (1-\lambda_B)\,\eta^*_i,
\ee
where $\widetilde{\eta}_i$ is now an independent Gaussian complex noise with the same properties (\ref{noisevars}) as the old $\eta^{\pr}_i$.

\subsection{Entanglement scaling performance}
\label{ssn: extrap-results}
\begin{figure}
\centering
\includegraphics[width=0.9\columnwidth]{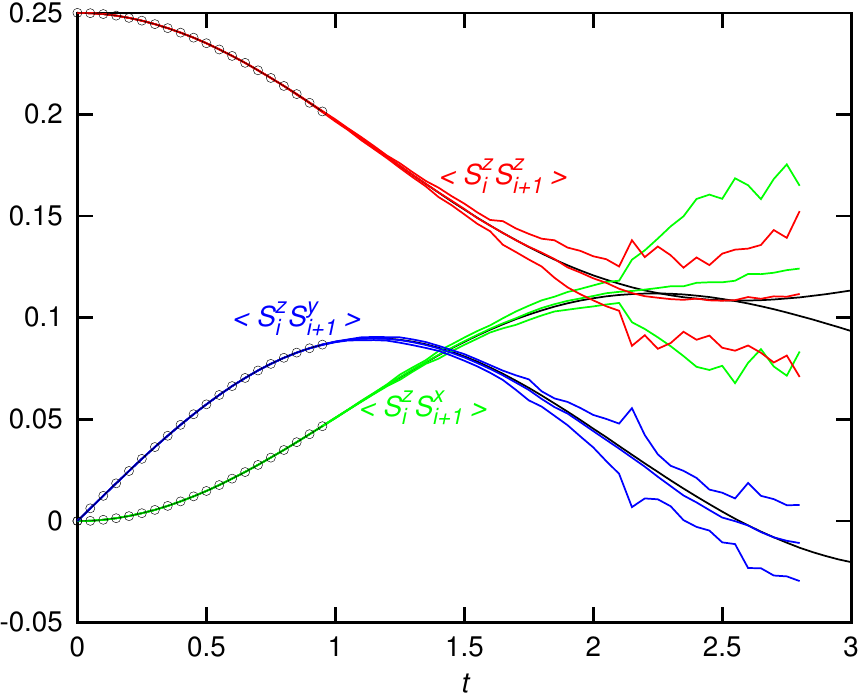}\vspace*{-0.3cm}
\caption{(color online).  
Predictions of the nearest neighbor correlations involving $S^z$ for the $h=0.5$ Ising quench on $N=10$ sites of a 1D Ising chain, as obtained using the entanglement scaling method (triple colored lines), and compared with known exact values (black lines). Black markers show the predictions available with $\lambda_A=1$ direct calculations.
\label{fig:Sxyz}\vspace*{-0.3cm}	
}
\end{figure}

\begin{figure}
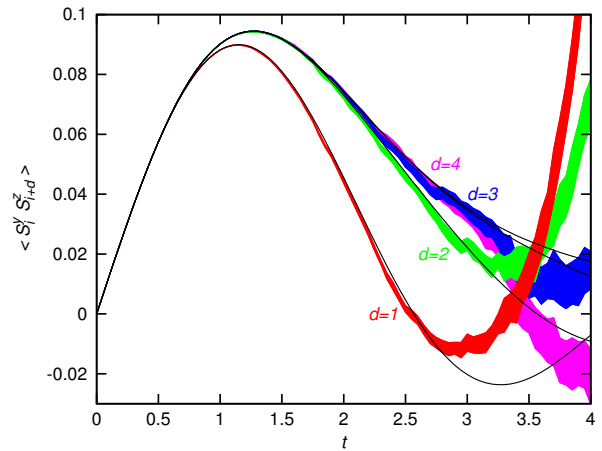

\centering
\includegraphics[width=0.9\columnwidth]{{{figure7}}}\vspace*{-0.3cm}
\caption{(color online).  Correlation between $\hat{S}^y$ and $\hat{S}^z$ spin components as a function of site-to-site distance $d$ for a $N=10$ site 1D lattice at the critical quench value $h=0.5$, as calculated using family A equations. Statistical uncertainty is shown as width of the color bars, while black lines show the exact results.  
\label{fig:corr}\vspace*{-0.3cm}
}
\end{figure}

Some predictions obtained with the fully-deployed entanglement scaling approach are shown in Figs.~\ref{fig:extrap}-\ref{fig:corr}. 
The first of these figures shows some detail of the procedure for the nearest neighbor correlation $\la\hat{S}^z_{i}\hat{S}^z_{i+1}\ra$ (see also Appendix~\ref{ssn: extrap-proc} and Fig.~\ref{fig:ll} for more). 

One can see that at longer times, the predictions of both families (green and yellow regions) are much closer to the true value than any of the magenta or cyan lines that were directly simulated. 
An obvious feature is that the family A prediction gives a much smaller statistical uncertainty than the family $B$ prediction. This is related to the longer range of data in $\lambda_A$ than in $\lambda_B$ that is available at a given time -- see also Fig.~\ref{fig:ll}. Hence, for the mean final estimate (blue, central line) we use the family A estimate. The uncertainty in our final prediction is taken to be the maximum of three values: the statistical uncertainty from the family A and family B predictions, as well as the absolute difference between the mean family A and family B predictions. The last value takes into account any systematics due to the extrapolations in $\lambda_m$ without needing to refer to any exact calculations, so that a reasonable uncertainty estimate can be obtained for large systems when no exact result is available.

Figure~\ref{fig:Sxyz} further compares the predictions and their uncertainty with the true values, which can still be calculated for $N=10$.
It shows that we obtain useful results until $t\approx2.8$, which is about three times longer than the plain approach of Sec.~\ref{ssn: quench results}. 
This is long enough to access the particularly important intermediate time regime of the problem \cite{Essler12-I} that occurs when $v_{\rm max} t \sim d$ for correlations between $d$th neighbours, where 
$v_{\rm max} = 2|J|{\rm min}[h,1]$ is the maximum propagation velocity. For our examples here, $v_{\rm max}=1$. This is also characteristic timescale for single-site decoherence in the system such that $|\la \hat{S} \ra|$ decays to its typical long-time value over this time. 

We see that the match in Fig.~\ref{fig:Sxyz} is well within the uncertainty reported, and in fact the uncertainty given is quite conservative. The limiting factor is the relatively poor performance of family B in comparison with family A. For this system, the Glauber-Sudarshan-like equations give results further from the full quantum value than the noiseless ones. To reduce the final uncertainty, one needs a substitute,  ``family C'', that gives results that deviate less 
while retaining a simple dependence on $\lambda_C$.

\begin{figure}
\centering
\includegraphics[width=\columnwidth]{{{figure5}}}\vspace*{-0.3cm}
\caption{(color online).  
Detail of entanglement scaling shown for the example of the
nearest neighbor correlation $\la\hat{S}^y_{i}\hat{S}^z_{i+1}\ra$ for the $h=0.5$ Ising quench on $N=10$ sites of a 1D Ising chain. Cyan/magenta sets of lines show the predictions obtained for different $\lambda_m$ values  with family A and family B equations, respectively. Green and yellow zones show the extrapolation to the full quantum values $\lambda_m=1$ obtained by each of the two methods,  respectively. Vertical width gives the statistical uncertainty. The blue triple lines give the final combined estimate and uncertainty. For comparison, the black markers show the predictions available with $\lambda_A=1$ direct calculations, while the black line shows the exact value. 
\label{fig:extrap}\vspace*{-0.3cm}	
}
\end{figure}

For example, Figure~\ref{fig:corr} shows the build-up of correlations at a range as time progresses, when calculated using family A data. One can follow the propagation of the disturbance created by the quench by observing the times at which the correlation values diverge from each other with subsequent $d$. It would be highly advantageous to have a second family with similar statistical uncertainty, so as to be able to continue resolve the difference between $d=2$ and $d=3$ in the final predictions. 

\subsection{Large systems}
\label{ssn: large}

\begin{figure}
\centering
\includegraphics[width=0.9\columnwidth]{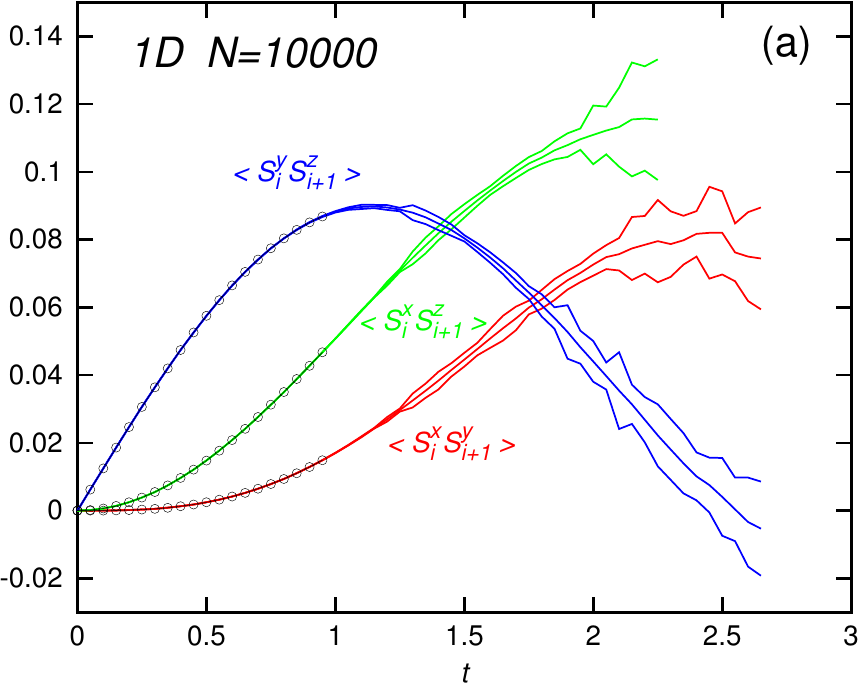}\\
\includegraphics[width=0.9\columnwidth]{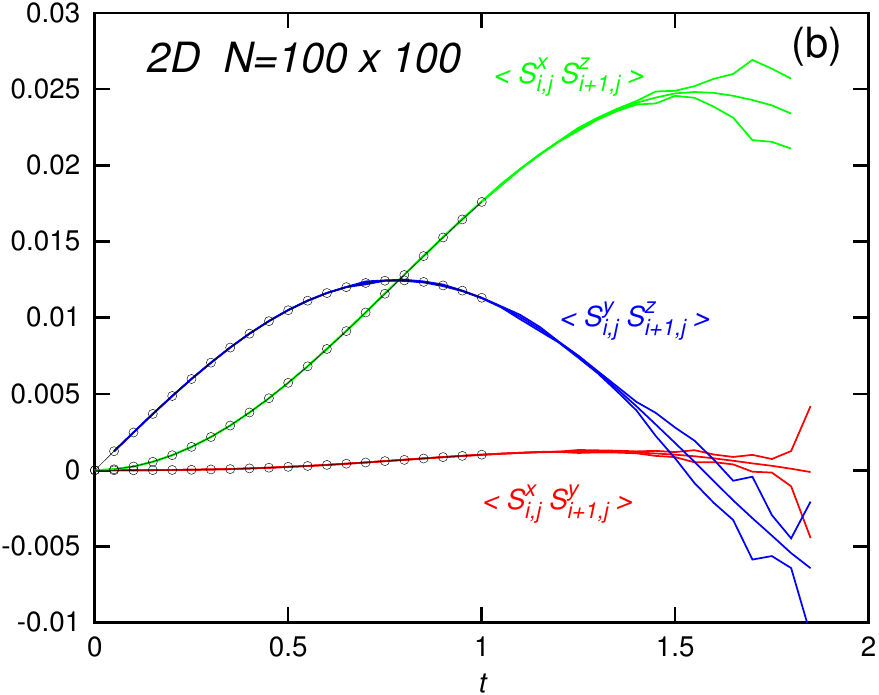}\vspace*{-0.3cm}
\caption{(color online).  Dynamics of large spin systems after a quench, showing  nearest-neighbor correlations between orthogonal spin directions. 
 Panel (a): A $N=10000$ 1D Ising spin chain quenched to $h=0.5$; Panel (b): A $100\times100$ spin lattice quenched to $h=0.1$. 
Triple lines show prediction and uncertainty obtained via the entanglement scaling method. Black markers show the predictions available with $\lambda=1$ direct calculations.
\label{fig:others}
}\vspace*{-0.3cm}
\end{figure}

In Fig.~\ref{fig:others} we show predictions for two very large systems (A spin chain with $N=10^4$ sites, and a 2D square lattice with $100\times100$ sites, inaccessible with direct calculation. Indeed, the full quantum dynamics of a 2D case of the size shown in Fig.~\ref{fig:others}(b) are presently
numerically intractable by any other currently available methods.  

Importantly, the lifetime for the 1D spin chain calculations shown in Fig.~\ref{fig:Sxyz} for $N=10$ is the same as for $N=10000$ in Fig.~\ref{fig:others}. This confirms that 
with these methods the simulation performance need not depend intractably on the system size. Naturally, being able to simulate $10^4$ spins would become especially useful for such cases as non-uniform systems or quenches, rather than the uniform test cases shown here. 




%

\section{Comments and Conclusions}
\label{sn: conclusions}

\subsection{Summary}
We have implemented a phase-space representation for spin systems based on the SU(2) coherent states and demonstrated that it can be used to simulate the full quantum real-time dynamics of large systems of interacting spins, giving correct results.

A direct application of the representation allowed us simulate the dynamics for significantly longer times than previous attempts using Schwinger bosons \cite{Ng11}, e.g. an improvement from $\sim 0.6 \hbar/J$ (in \cite{Ng11}) to  $1.1\hbar/J$ for Ising chains after a transverse field quench to the critical value of $h=0.5$. By using the entanglement scaling technique, we have been able to extend simulated times further, to times of up to $2.8\hbar/J$, which is long enough to observe the main decoherence effects and the propagation of correlations.
Furthermore, initial states are compact so that these representations now exhibit good scaling with system size --- the times achievable do not depend on the number of spins in the system, apart from computer resource limitations which scale only linearly with the number of inter-spin coupling terms. This allows one to access really large systems that are not directly accessible by other methods, such as the $10^4$ spins calculations in one and two dimensions  demonstrated here.

\subsection{Outlook}
This work is a first application of the entanglement scaling approach \cite{Deuar09} beyond BEC collisions. 
Avenues for further improvement of simulated times $t_{\rm sim}$ include diffusion stochastic gauges \cite{Deuar06b} to reduce diffusion of trajectories into badly-normalized regions of phase-space such as ${\rm Im}[R]$, or a combination of drift and diffusion gauges of the kind presented by Dowling\etal \cite{Dowling07} with Metropolis sampling of the resulting real weights. For application of entanglement scaling, stabilization of the equations may be useful by the use of just drift gauges \cite{Deuar02,Drummond04} or the methods presented in Perret\etal \cite{Perret11}. Finding a third family of equations, ``family C'', that more closely matches the full evolution at $\lambda_m=0$ than family B, would strongly improve the precision of the final estimates. 

Perhaps the most promising avenue to consider  is to build a different kind of kernel that is more closely suited to the natural states of the Hamiltonian (\ref{eq: Hamiltonian-general}), especially some variety that builds nearest-neighbor correlations into the basis. To this end, the conjectures at the end of Sec.~\ref{ssn: originpoles} are points to remember when formulating new kinds of phase-space descriptions.

Within the existing time limitations, there is a range of problems for which short-time spin dynamics can tell us a lot. This includes quantum quenches in general, the study  of critical behavior and the pinpointing of phase transitions by analysis of the Loschmidt echo \cite{Rossini07}. The coherence properties of a system can be investigated with echo sequences of external forcing parameters \cite{Raitzsch08,Younge09,Dowling05,Wuster10}, something that is especially useful for lossy systems because it alleviates the need for evolution over long times.  The representations developed here can be used to simulate such situations without imposing approximations or projections onto the Hamiltonian, especially in 2D and 3D systems -- something for which efficient methods have been lacking.

\acknowledgments{
We are grateful to Peter Drummond and Joanna Pietraszewicz for helpful discussions. We also acknowledge financial support from the EU Marie Curie European Reintegration Grant PERG06-GA-2009-256291, the Polish Government grant 1697/7PRUE/2010/7 as well as from NSERC.
This work was made possible by the facilities of the Shared Hierarchical Academic Research Computing Network (SHARCNET:www.sharcnet.ca) and Compute/Calcul Canada.
}


\bibliographystyle{prsty}
\bibliography{su2}

\appendix

\section{Remapping the variables onto a seamless space}
\label{sn: new vars}

\subsection{Polar and Boolean variables}
To allow the representation of the $|\!\!\uparrow\rangle$, $|\!\downarrow\rangle$ states polarized in the $\vec{z}$ direction, and avoid the stiffness of the equations near these points in phase space, we make a change of variables that prevents such infinite values. 
While the surface of the Bloch sphere cannot be seamlessly mapped onto a plane, hemispheres are easily treated. We make a transformation similar to a polar projection centered on the nearest pole, and introduce a Boolean variable $s$ that keeps track of which pole is being used for a given trajectory. 

For a state $|z\ra$ we implement the following transformation to a complex variable $y$:
\be
 y =  \left\{ \ba{ccc}  e^{-z},&  {\rm\ and\ } s=+1 &\text{  if    }  \mathbb{R}[z] >0  
\\
   \\   \\  e^{z^{\ast}},& {\rm\ and\ } s=-1 &\text{  if    }  \mathbb{R}[z]  \leq 0   
\ea \right.,
\ee
where the variable $s={\rm sign}\left(\mathbb{R}[z]\right)$ tells us which Bloch hemisphere we are in. 
Under this parametrization, the variable $y$ never leaves the unit circle $|y|\le1$. The extreme spin values of $S_z=\pm\frac{1}{2}$ are now at the well-behaved $y=0$ point, with $s=\pm1$.

Since the branch cut in this parametrization lies on the far pole, the trajectories can never go near this singular region so long as we make sure to change the parametrization whenever the trajectory crosses the ``equator''. This is implemented by checking at the end of each time-step whether $|y|$ has crossed outside the unit circle. If it has, we carry out 
\be\label{coordchange}
y \to \frac{1}{y^*}\qquad{\rm and}\qquad s \to -s.
\ee
When time steps are small, there is then no risk of approaching the far, pathological, pole. The evolution near the equator of the Bloch sphere is gradual, although swapping between projections occurs.

For the many-mode system, we need separate variables $y_i$, $y^{\pr}_i$, $s_i$, and $s^{\pr}_i$ for each spin. 
The SDEs (\ref{dzdt})-(\ref{dzidt}) take on slightly different forms depending on which hemispheres the bra and ket components of the kernel lie. In terms of only the new variables, they are:
\be
\f{dy_i}{dt}   = \left\{   \ba{cl}   \ba{c} -i y_i \left\{ S^z_{i-1} + S^z_{i+1} \right\} + \f{ih}{2}  ( 1- y_i^2)  \vspace*{0.3em}\\
  - y_i \left[  \eta_i + i\eta_{i-1}^{\ast}  \right ] \ea  &  \text{ if } s_i=+1 \\
& \\
 \ba{c} -i y_i \left\{ S^{z\ast}_{i-1} + S^{z\ast}_{i+1} \right\} - \f{ih}{2}  ( 1- y_i^2)\vspace*{0.3em}\\
  + y_i \left[  \eta_i^* -i \eta_{i-1}  \right ] \ea  &  \text{ if } s_i=-1 \ea\right.\label{dydt}
\ee
and 
\be
\f{dy^{\pr}_i}{dt}   = \left\{   \ba{cl}   \ba{c} i y_i^{\pr} \left\{ S^z_{i-1} + S^z_{i+1} \right\} - \f{ih}{2}  ( 1- y_i^{\pr 2})  \vspace*{0.3em}\\
  - y^{\pr}_i \left[  \eta^{\pr}_i -i \eta_{i-1}^{\pr \ast}  \right ] \ea  &  \text{ if } s^{\pr}_i=+1 \\
& \\
 \ba{c} i y^{\pr}_i \left\{ S_{i-1}^{z\ast} + S_{i+1}^{z\ast} \right\} + \f{ih}{2}  ( 1- y^{\pr 2}_i)\vspace*{0.3em}\\
  + y^{\pr}_i \left[  \eta^{\pr \ast}_i + i\eta^{\pr}_{i-1}  \right ] \ea  &  \text{ if } s^{\pr}_i=-1 \ea\right.\label{dyidt}
\ee
where the $S^z$ estimator in terms of the new variables is:
\bea
2 S^z_i = \tanh R & = & \left\{  \ba{cl}  \f{1 - y_iy^{\pr}_i }{1 + y_i y^{\pr}_i }  &\text{if } s_i=+1, s'_i=+1  \\
 \f{y_i^{\pr \ast}  - y_i } {y^{\pr \ast}_i + y_i } &\text{if } s_i=+1, s'_i=-1  \\
 \f{y^{\ast}_i  - y_i^{\pr} } {y^{\ast}_i + y^{\pr} } &\text{if } s_i=-1, s'_i=+1  \\
\f{ y^{\ast}_i y^{\pr \ast}_i  -1  } { 1+ y^{\ast}_i y^{\pr \ast}_i   }&\text{if } s_i=-1, s'_i=-1
\ea \right.\label{4cases}
\eea

\subsection{Simulation termination}
\label{sn:spikes}
\begin{figure}
\centering
\includegraphics[width=\columnwidth]{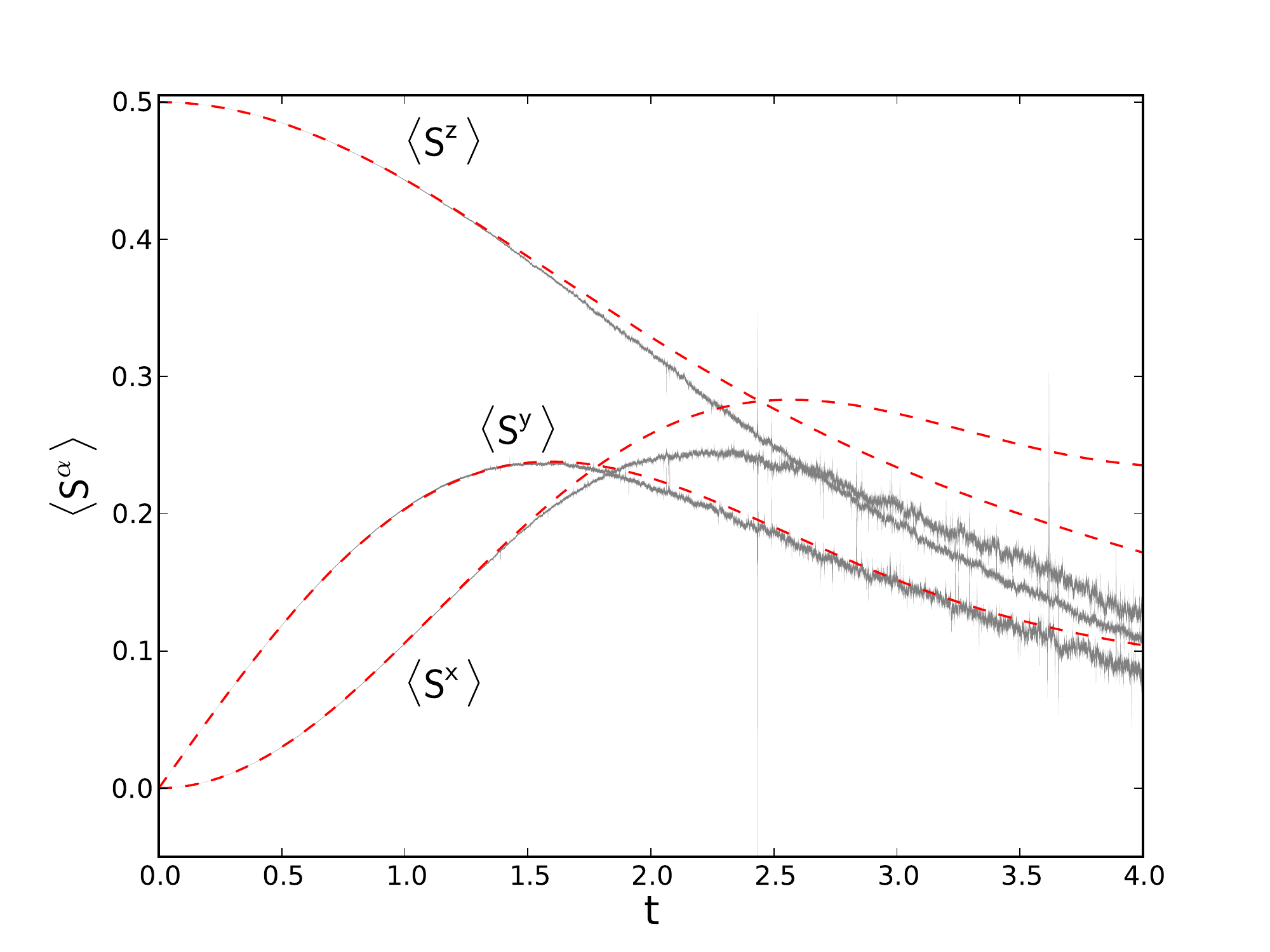}
\vspace*{-0.3cm}
\caption{(color online).  Spin components: $\la\hat{S}^x \ra, \la\hat{S}^y \ra , \la\hat{S}^z \ra$ (bottom to top) vs. time 
 for the ten-site Ising spin chain with transverse quench from $h=0$ to $h=0.5$.  Results of simulations using the seamless equations with $y,s$ variables. Here, 
we do not use the criterion (\ref{tanshing}) to stop the simulation, so as to show the bare behavior. $\mc{N}=10^4$ trajectories, $\mc{B}=100$ bins. Error bars and exact results are shown. The criterion (\ref{tanshing}) to stop the evolution 
is achieved at $t_{\rm sim} \approx 1.1$. 
\label{fig: zt0}}\vspace*{-0.3cm}
\end{figure}

At times $t\gtrsim t_{\rm sim}$, as shown in Fig.~\ref{fig:hdep}, a pronounced spiking behavior is seen in observable means. It is caused by  approaches to the $R=\pm i\pi/2$ poles described in Sec.~\ref{ssn:limitations}. Spikes are a warning sign that poor sampling of the distribution may be occurring \cite{Gilchrist97}, so one should disregard the simulation for times after its onset.  With the original $z,z'$ variable equations that have stiff behavior, spikes also lead immediately to numerical inaccuracy and overflow, so that ensemble averages of the estimators also overflow and any actual spiking / systematic error is hidden from view. A similar behavior was seen in positive-P simulations of boson fields \cite{Deuar06}. The seamless variables $y,y'$ are less stiff so that overflow does not occur and the bare spiking behavior can in principle be seen. An example is shown in Fig.~\ref{fig: zt0}. 
The results shown in other figures disregard evolution after the appearance of the first spike. We detect spikes by checking whether 
\be\label{tanshing}
|S^z| = \frac{1}{2}|\tanh R_i|>1/\epsilon,
\ee
for any trajectory at any site $i$, where we choose $\epsilon=0.04$.

\section{Entanglement scaling procedure}
\label{ssn: extrap-proc}

\begin{figure}
\centering
\includegraphics[width=0.9\columnwidth]{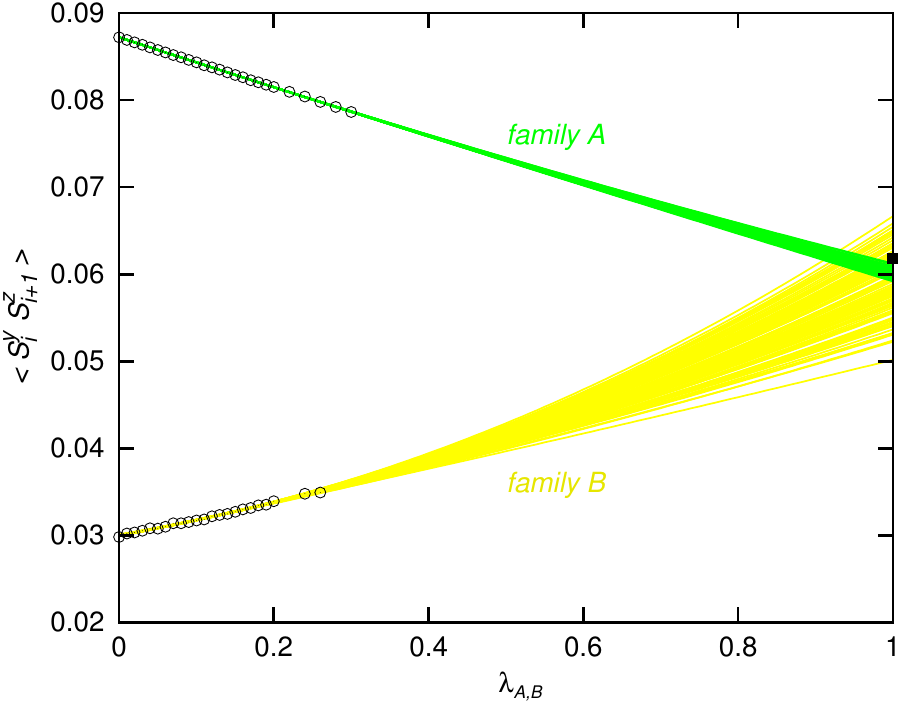}\vspace*{-0.3cm}
\caption{(color online).  Extrapolation details for the
nearest neighbor correlation $\la\hat{S}^y_{i}\hat{S}^z_{i+1}\ra$ for the data of Fig.~\ref{fig:extrap} at $t=1.8$. The black circles show observable estimates $O_m(\lambda_m)$ obtained with family A and B equations. 
The green and yellow lines show the $j=1,\dots \mathcal{S}=100$ ensemble of quadratic fits $f_m^{(j)}(\lambda_m)$ obtained for family A and B, respectively, with the synthetic data sets $O_m^{(j)}(\lambda_m)$. The black square is the exact quantum value obtained through exact diagonalization. 
\label{fig:ll}}\vspace*{-0.3cm}
\end{figure}

The technique is described in detail in \cite{Deuar09}. We proceed as follows:
\begin{enumerate}
\item For an observable of interest $\hat{O}$, we generate observable estimates $O_m(t,\lambda_m)\pm\Delta O_m(t,\lambda_m)$ for a sequence of $\lambda_m$ values, for each method $m$. Here, $m=\{A,B\}$. We expect that at long times $t>t_{\rm sim}^m(\lambda_m)$, the data are missing (here, due to rejection because of the onset of spiking). Seen in Figs.~\ref{fig:extrap} and~\ref{fig:ll}. 
\item We use the available $\lambda_m$ ranges of data to extrapolate to the full quantum predictions $Q_m(t)$ at $\lambda_m=1$. 
\item We estimate the statistical uncertainty of these extrapolations $\Delta Q_m(t)$. 
\item The final best estimate $Q(t)$ is taken to be the prediction $Q_m(t)$ with the smallest uncertainty among the $\Delta Q_m(t)$.
\item The final uncertainty $\Delta Q(t)$ is taken to be the maximum among all statistical uncertainties $\Delta Q_m(t)$ and discrepancies $|Q_m(t)-Q(t)|$. The latter takes into account systematics due to poor fits without needing to refer to any exact calculations. 
\end{enumerate}

Two or more families $m$ are used to provide a check on each other's accuracy. For this to work, they must make independent estimates of $Q_m(t)$. Since we are in principle free to choose the functional form by which $\lambda_m$ enters the evolution equations, estimates will only be independent when the $\lambda=0$ starting points differ to a statistically significant degree. 

There is also the matter of choosing fitting functions in the $\lambda_m$. In principle they are unknown \textit{a priori}. In practice, complicated dependences on $\lambda_m$ are unacceptable because the extrapolation would become ill-conditioned due to porly constrained fitting parameters.
Scaling the noise variance with $\lambda_m$ tends to give near-linear dependence, when the result at $\lambda_m=0$ is a noiseless set of equations. 
We try polynomials up to third order as our fitting functions $f_m(t,\lambda_m)$ in $\lambda_m$. 
In almost all cases quadratic fits give the best results -- linear fits tend to  disagree between methods $m$ by more than statistical uncertainty because the dependence is too simple, while cubic fits are usually ill-conditioned and give huge uncertainties. 

Uncertainty estimates for extrapolations can be found by various means \cite{NumericalRecipes}. 
One relatively straightforward method is to generate a set of synthetic data sets where deviations from the fit are randomized. To do this, we 
calculate the rms deviation from the fit $\mc{R}_m(t)^2 = \frac{1}{N_m(t)}\sum_{\lambda_m} (O_m(t,\lambda_m) - f_m(t,\lambda_m))^2$, and add random Gaussian noise having this standard deviation to the original data. $N_m(t)$ is the number of $\lambda_m$ values used. This gives an ensemble of data sets labeled by $j = 1,\dots,\mc{S}$, with values 
\be
O_m^{(j)}(t,\lambda_m) = O_m(t,\lambda_m) + \mc{R}^{\rm cap}_m(t,\lambda_m)\xi^{(j)}_m(t,\lambda_m),
\ee
where the $\xi^{(j)}_m(t,\lambda_m)$ are independent Gaussian random variables with mean zero and variance unity. In practice we use a deviation that is capped from below $\mc{R}^{\rm cap}_m(t,\lambda_m) = {\rm max}[\Delta O_m(t,\lambda_m),\mc{R}_m(t)]$ to not be smaller than the statistical uncertainty in the data points. 

Having these sets, an extrapolation $Q_m^{(j)}(t)$ is made with each one. The uncertainty in our final estimates $\Delta Q_m(t)$ is based on the distribution of $Q^{(j)}_m(t)$. This need not be Gaussian, so instead of using standard deviations we consider percentiles (68.3\% in the figures). We use $\mc{S}=100$ such synthetic data sets in each case. 


\section{Simulation time for nonzero $h$}
\label{sn:h}

Consider the equations (\ref{dydt}) for the ``seamless'' $y$ variables, 
and let us stay now generally in the $s=s'=+1$ projection, since what follows is very approximate and this simplification is sufficient to obtain the observed scaling (\ref{hdeplo}) and (\ref{hdephi}).  
The poles correspond to the denominator in (\ref{4cases}) going to zero. i.e. for the $s=s'$ case, when $yy'=-1$. Since 
$|y|\le1$, this means
\be
y' = -y, \qquad{\rm and}\qquad |y|=|y'|=1.
\ee
is the location of the poles. I.e. $y$ and $y'$ lie opposite each other on the unit circle. 

To estimate when this can occur, consider that $y$ and $y'^*$ start out equal, and have a similar evolution that differs by some random noise. Hence, the variance of the distance $|y-y'|$ is of the same order as the variance of $|y|$. 
Poles can occur only when $|y-y'|=2$, so we expect approaches to the poles to begin when the variance of $y$ is of the order of half (then  $\pm2\sigma$ outliers are separated by a distance of $\approx 2$).
We will estimate $t_{\rm sim}$ by looking for the time when 
\be\label{crit}
{\rm var}[|y|] = \la|y|^2\ra-|\la y\ra|^2 = \Delta_y^2
\ee
where $\Delta_y$ is a constant $\mc{O}(1/2)$.

Let us look at the evolution of $y$. Initially all the $S^{z}_{i-1}$ and $S^z_{i+1}$ contributions are negligible because they are multiplied by $y\approx0$. If we ignore them, the sites decouple, and one 
has some hope of a simple analysis, so let us proceed in that way. The noises $\eta_{\la i,j\ra}$ can be collected together into one larger noise, and the approximate equation is:
\be\label{dydt0}
\frac{dy}{dt} = \frac{ih}{2}(1-y^2) +y\sqrt{C}\eta(t).
\ee
where $C$ is the number of connections per site. For example, $C=2d$ in $d$-dimensional square lattices. $\eta$ has the same statistical properties (\ref{noisevars}) as one of the $\eta_i$. 

Initially, $y\approx0$, and (\ref{dydt0}) leads to 
\be\label{approxylint}
y(t)\approx i\frac{ht}{2}.
\ee
so that the trajectories move upwards towards $y\to i$, while starting to acquire fluctuations. 
There are two extreme possibilities: either the trajectories all move up towards the unit circle without acquiring much noise along the way (large $h$), or their average stays small while the outliers approach the unit circle (small $h$). Let us, for now, ignore also the nonlinearity that occurs when $y\sim\mc{O}(1)$, and consider the early-time equation
\be\label{dydt1}
\frac{dy}{dt} = \frac{ih}{2} +y\sqrt{C}\eta(t)
\ee 
which can be solved. For $y(0)=0$, it is:
\be\label{soly1}
y(t) = \frac{ih}{2}\ e^{-\sqrt{C}f(t)}\ \int_0^t e^{\sqrt{C}f(s)}ds,
\ee
where the function
\be\label{fdef}
f(t) = \int_0^t \eta(s) ds = f_R(t)+if_I(t)
\ee
is an integrated noise that has the following properties:
\be\label{fprop}
\la f(t)\ra = 0,\qquad\la f(t)f(t') \ra = 0,\qquad \la f^*(t)f(t') \ra = |t-t'|
\ee

and $f_R$ and $f_I$ are the real and imaginary parts, respectively. These are independent and have equal variances of $|t-t'|/2$.

To proceed, we will need the following results \cite{DeuarPhD}, valid for real Gaussian random variables $\xi$ of variance 1 and zero mean:
\be\label{gausresult}
\la e^{\sigma\xi} \ra = e^{\sigma^2/2}, \qquad \la e^{i\sigma\xi} \ra = e^{-\sigma^2/2}.
\ee

Let us now evaluate the variance of $|y|$. The expression (\ref{soly1}) for $y$ can be grouped according to independent noises in the exponential, such that 
\be\label{yform}
y = \frac{ih}{2}\int_0^t ds\ e^{\sqrt{C}(f_R(t)-f_R(s))}\ e^{i\sqrt{C}(f_I(t)-f_I(s))}.
\ee
Each factor with independent noises can be evaluated independently, so 
\be
\la y\ra = \frac{ih}{2}\int_0^t ds \la e^{\sqrt{C}(f_R(t)-f_R(s))}\ra\la e^{i\sqrt{C}(f_I(t)-f_I(s))}\ra.
\ee
The noise difference is $f(t)-f(s) =  \int_s^t \eta(s') ds'$ and its real and imaginary parts has a variance of $|t-s|/2$. Then, using (\ref{gausresult}) we obtain 
\be
\la y\ra = \frac{iht}{2}.
\ee
A similar but slightly more lengthy procedure using the substitution (\ref{yform}) gives 
%
$\la|y|^2\ra =   \frac{h^2t}{4C}\left(e^{Ct}-1\right)$ so that
\be\label{vary1}
{\rm var}[|y|] = \frac{h^2t}{4C}\left[e^{Ct}-1-Ct\right],
\ee
to be compared with our variance criterion (\ref{crit}).

The approximate deterministic evolution (\ref{approxylint}) is valid as long as $\la y\ra$ remains small, i.e. for times  $\lesssim t_0=2/h$. When $h$ is small, $t_0\gg1$, and the variance at this time is 
$\approx e^{2C/h}(h/2C)\gg1$. Hence, our variance criterion (\ref{crit}) for $t_{\rm sim}$ is exceeded while our assumptions hold. 
Under the $2/h\gg1$ assumption,
$
{\rm var}[|y|] \approx \frac{h^2t}{4C}e^{Ct},
$
so that \textbf{for small $h$} (\ref{crit}) gives
\be\label{hdeplowh}
t_{\rm sim} \approx \frac{2}{C}\log\left(\frac{2\Delta_y\sqrt{C}}{h}\right) - \frac{1}{C}\log t_{\rm sim} \approx \frac{2}{C}\log\left(\frac{2\Delta_y\sqrt{C}}{h}\right).
\ee
The $\log t$ term is negligible for small enough $h$, so the observed scaling behavior (\ref{hdeplo}) is recovered. 
A comparison with the data of Fig~\ref{fig:hdep} gives a match for $\Delta_y\approx1/4$. 

Different behavior occurs when $h$ is large. In this case, by the ``large-$y$'' time of $t_0$,  (\ref{vary1}) gives 
\be\label{vart0}
{\rm var}[|y|(t_0)]\approx \frac{C}{h} \ll1
\ee
so that by the time the linear drift approximation (\ref{approxylint}) used to obtain (\ref{vary1}) breaks down, the variance is still small and the $yy'=-1$ poles have not been approached. 

At later times, 
if we continue to ignore the Ising drift terms $\sim yS^z$, then upon reaching $y\approx i$, we make the coordinate change (\ref{coordchange}) to obtain $s\to-1$, and $y\to\approx-i$, soon followed by also a complementary flip in $s'$ and $y'$, since the variance of the trajectories is small. The evolution then continues to drift upwards according to $\dot{y}\approx ihy/2$ until we again reach $y\approx +i$, and so on. This basically corresponds to precession invoked by the strong transverse field along the $\vec{x}$ axis. For large $h$, many such periods will occur before $t_{\rm sim}$ is reached. Let us make a gross approximation that on average the value of $|y|$ is $\wb{y}$ in the time period $t_0<t<t_{\rm sim}$, expecting $\wb{y}$ to be $\mc{O}(1)$.  Then an approximate equation of motion is 
\be\label{dydt2}
\frac{dy}{dt} = {\rm\ deterministic\ terms } +\wb{y}\sqrt{C}\eta(t).
\ee 
The equation of motion for the variance, on the other hand, is (via the Ito calculus)
\be
\frac{d}{dt}{\rm var}[|y|] = \frac{C|\wb{y}|^2}{2} + \left(\la y^*\frac{dy}{dt}\ra - \la\frac{dy}{dt}\ra\la y^*\ra\right) + {\rm c.c.}
\ee
The covariances on the right hand side can have a complicated dependence on $y$, but in the spirit of simplifying down to the bare essentials, let us omit them. With that, we find
\bea
{\rm var}[|y(t)|] &\approx& {\rm var}[|y(t_0)|] + \frac{C|\wb{y}|^2}{2}(t-t_0)\\
&\approx& C\left[ \frac{1}{h} + \frac{|\wb{y}|^2}{2}\left(t-\frac{2}{h}\right)\right],\label{vary2}
\eea
using also (\ref{vart0}). Applying the criterion (\ref{crit}) we obtain the following estimate \textbf{at large $h$}:
\be\label{hdephighh}
t_{\rm sim} \approx \frac{2\Delta_y^2}{C|\wb{y}|^2} + \frac{2}{h}\left(1-\frac{1}{|\wb{y}|^2}\right),
\ee
the general behavior being (\ref{hdephi}). A comparison with the data of Fig~\ref{fig:hdep} gives a match for $c_1\approx 0.8$ and $c_2\approx0.3$, i.e. $\wb{y}\approx1.1$ and 
$\Delta_y\approx0.7$.  


\end{document}